\documentclass{article}

\usepackage[final]{aaj2026}
\usepackage[numbers]{natbib}

\usepackage[utf8]{inputenc} 
\usepackage[T1]{fontenc}    
\usepackage{hyperref}       
\usepackage{url}            
\usepackage{booktabs}       
\usepackage{amsfonts}       
\usepackage{nicefrac}       
\usepackage{microtype}      
\usepackage{xcolor}         

\usepackage{graphicx}
\usepackage{subcaption}
\usepackage{float}
\usepackage{orcidlink}

\graphicspath{{figuras/}}

\title{Beyond Vector Search: Comparing Classical RAG with Hybrid GraphRAG for Climate Science Q\&A}

\author{%
  Daniel Naiff\orcidlink{0000-0000-0000-0000} \\
  Universidade Federal do Pará\\
  Belém, PA, Brazil \\
  \texttt{danielnaiff344@gmail.com} \\
\And
  Ronnie Alves\orcidlink{0000-0000-0000-0000} \\
  Instituto Tecnológico Vale\\
  Belém, PA, Brazil \\
  \texttt{ronnie.alves@itv.org} \\
\And
  Gustavo Pinto\orcidlink{0000-0003-4900-6369} \\
  Universidade Federal do Pará\\
  Belém, PA, Brazil \\
  \texttt{gpinto@ufpa.br} \\
}

\begin{document}

\maketitle

\begin{abstract}
  Traditional Retrieval-Augmented Generation (RAG) systems treat documents in isolation, failing to capture hierarchical relationships between concepts in complex scientific corpora. This limitation compromises answer quality in specialized domains such as climatology, where conceptual dependencies frequently traverse multiple articles. We propose a hybrid architecture that integrates vector search with GraphRAG, Leiden community detection, and cross-encoder re-ranking, achieving gains of 160\% in contextual relevance and 177\% in contextual recall compared to classical RAG. These results demonstrate that unifying local and global retrieval significantly outperforms text-span isolation, paving the way for more effective question-answering systems over dispersed scientific literature.
\end{abstract}

%
%
%

\section{Introduction}

The Amazon plays a fundamental role in the global climate balance, acting as a carbon sink and as a regulator of precipitation regimes in South America. This prominence has driven a substantial growth in the production of scientific articles about the biome. Although positive, this volume of publications creates a challenge for question-answering systems: extracting relevant knowledge from dispersed documents.

Retrieval-Augmented Generation (RAG) systems have emerged as a promising solution by combining Large Language Models (LLMs) with vector search. However, the traditional vector-based approach treats documents in isolation, without capturing hierarchical relationships between concepts. Recent work proposes evolutions of this architecture, such as those based on knowledge graphs (GraphRAG), which show potential to overcome such limitations by extracting entities, relations, and knowledge communities.

This paper compares traditional RAG with GraphRAG~\cite{edge2024local}, one of its most specialized branches. Both approaches were applied to a question-answering system whose database consists of scientific articles produced by a research group at Instituto Tecnológico Vale (ITV) (more details in Section~\ref{sec:arquitetura}).
For the experiments, we curated a dataset of 22 questions along with the answers expected by the team of researchers who use the system. This dataset was evaluated with 7 LLMs of different sizes and architectures, using LLM-as-a-Judge metrics such as correctness, answer relevance, and faithfulness. The results indicate that the hybrid approach significantly outperforms traditional vector RAG, with gains of 160\% in contextual relevance and 177\% in contextual recall.


\section{RAG and GraphRAG Architectures}

The RAG architecture was proposed by Lewis et al.~\cite{lewis2020retrieval}, bringing together language models and information retrieval. In classical RAG, documents are indexed as vector embeddings; at query time, the chunks that are semantically most similar to the question compose the context of the generator. Although effective, this approach fails to model explicit relations between concepts or isolated documents, which motivates structural extensions.
GraphRAG, in turn, proposed by Edge et al.~\cite{edge2024local}, overcomes this limitation by integrating knowledge graphs. The approach extracts textual entities and relations via an LLM, applying the Leiden algorithm to detect hierarchical communities and to generate multi-level summarizations. This allows the system to explore global and distributed connections in the corpus. The benefit of this structural representativeness is supported by Sequeda et al.~\cite{sequeda2024benchmark}, who show that zero-shot queries over raw data reach only 16\% accuracy, whereas introducing a semantic layer based on knowledge graphs raises this metric to 54\%.

These findings show that the reasoning of LLMs is limited without explicit connections. In the analysis of the ITV scientific collection, the need to correlate cross-cutting topics replicates this challenge. Therefore, the choice of GraphRAG tends to be justified by going beyond the geometric similarity of keywords, using the graph as a high-precision conceptual map to guide answer generation.

\section{Proposed Architectures}\label{sec:arquitetura}

This section describes the two architectures developed for the intelligent chatbot specialized in querying and synthesizing scientific articles of the partner institution. The first version was implemented based on the classical RAG approach. The second version improves the initial architecture by incorporating GraphRAG techniques, aiming to enrich the retrieved context, in addition to injecting the article title into each chunk generated during the vector indexing process.

The choice of a graph-based approach is justified by the need to go beyond the isolated geometric similarity of keywords of traditional vector search, explicitly mapping the knowledge correlated across multiple scientific documents on the climatology of the Amazon. In this structured scenario, the nodes (or vertices) of the graph represent the textual entities extracted from the documents (such as climate concepts, geographic locations, methodologies, authors, and institutions), while the edges (or links) map the semantic relations and explicit connections that exist between these entities (as, for example, the impact of a meteorological variable on a specific region or the co-authorship of a study). This structured modeling allows the community detection algorithm to organize knowledge in a macroscopic and hierarchical fashion, mitigating the isolation of textual passages.

The proposed system integrates: (i) an ingestion pipeline that converts PDFs, extracts metadata, translates content into Portuguese, generates chunks and embeddings, and builds a knowledge graph with hierarchical community detection and bottom-up summarization; and (ii) a query mechanism that unifies vector search and community summaries in a re-ranking process with a multilingual cross-encoder, giving equal weight to both sources of information.
As an architectural and project design guideline, all technological solutions and models employed in the ecosystem of the proposed system were based on open-source tools and technologies, ensuring reproducibility, autonomy, and control over the privacy and local processing of the scientific data.

\subsection{Overall Architecture, v1}

\begin{figure}[!ht]
    \centering
    \includegraphics[width=0.7\textwidth, trim=0mm 100mm 0mm 100mm, clip]{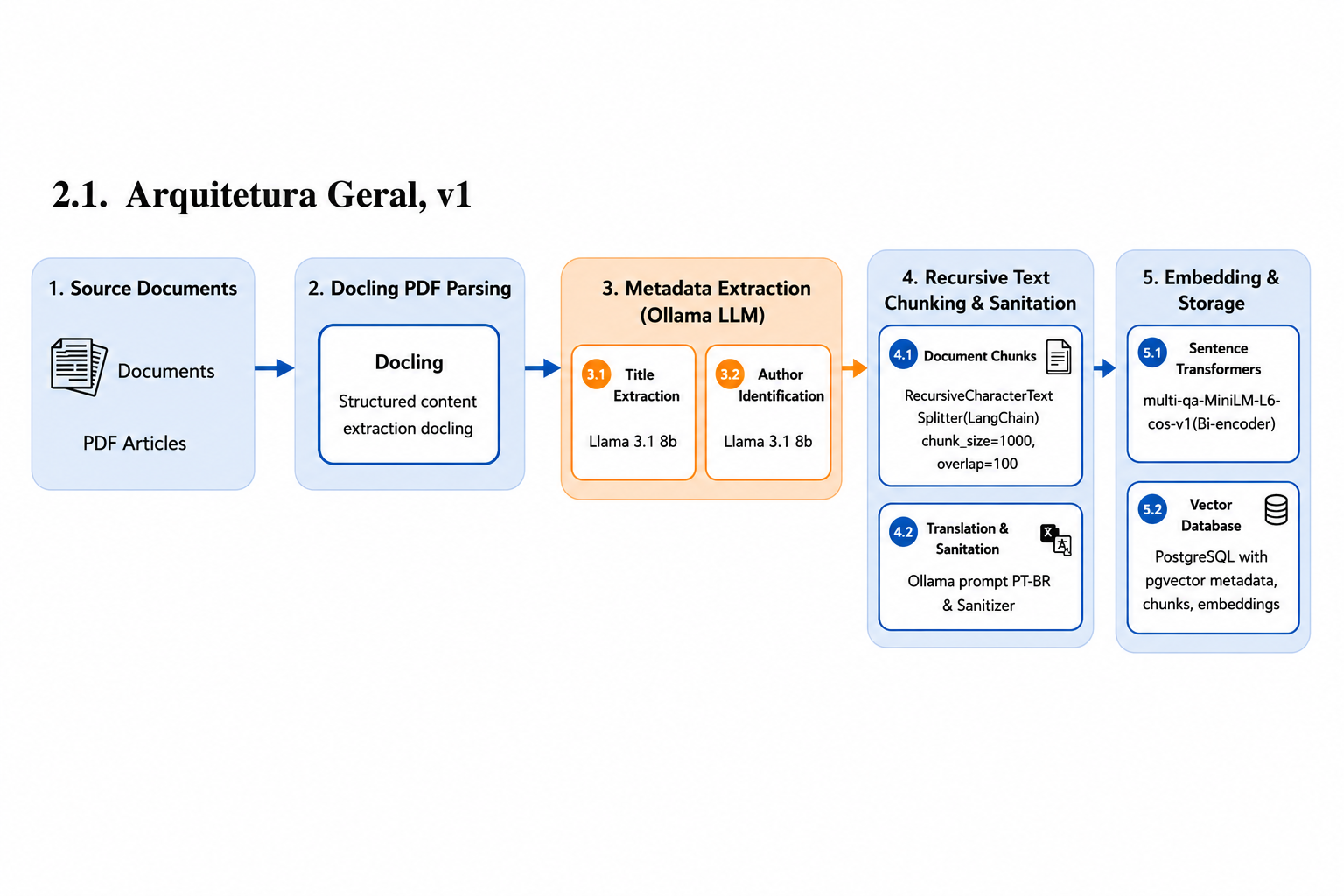}
    \caption{Ingestion pipeline of the first version of the chatbot.}
    \label{fig:iv1}
\end{figure}

Figure~\ref{fig:iv1} illustrates the ingestion pipeline of the first version of the system. This pipeline is responsible for processing the PDF documents and for building the vector knowledge base.

The flow starts with step 1 (\textit{Source Documents}), containing the articles in PDF format. Next, in step 2 (\textit{Docling PDF Parsing}), the structured extraction of the content of the PDF files takes place through the \texttt{Docling} library\footnote{\url{https://github.com/DS4SD/docling}}. In step 3 (\textit{Metadata Extraction (Ollama LLM)}), the first page of each article is processed for the automatic extraction of the article title and author information, along with other metadata. For this extraction, specific prompts are sent to the \texttt{llama 3.1 8b} model via Ollama, with the mapping of the title guided by sub-step 3.1 (\textit{Title Extraction}).

Subsequently, in step 4 (\textit{Recursive Text Chunking \& Sanitation}), the full text of the document is segmented in sub-step 4.1 (\textit{Document Chunks}) through the recursive chunking technique\footnote{Implemented by the \href{https://python.langchain.com/api_reference/text_splitters/character/langchain_text_splitters.character.RecursiveCharacterTextSplitter.html}{\texttt{RecursiveCharacterTextSplitter}} class, from the \texttt{LangChain} framework.}. Recursive chunking works hierarchically, using a list of default separators --- starting with paragraph breaks (``\texttt{\textbackslash n\textbackslash n}''), followed by simple line breaks (``\texttt{\textbackslash n}''), whitespace (`` ''), and finally empty characters (``'') --- splitting the text recursively whenever a block exceeds the configured maximum limit, which ensures the preservation of the structural integrity of complete sentences and paragraphs. After some experiments, the values of 1,000 characters and an overlap of 100 characters were defined. This means that each generated fragment contains at most 1,000 characters and, to avoid the loss of context at the cut boundaries, the last 100 characters of a block are replicated at the beginning of the following block, maintaining the semantic continuity between them. After the segmentation process, in sub-step 4.2 (\textit{Translation \& Sanitation}), each generated chunk goes through a process of sanitation and automatic translation into Portuguese. This is done because the linguistic alignment between the database and the user queries optimizes retrieval and ensures that the embeddings capture the semantic relations in the native language of the final system more precisely, reducing the cost and processing time at run time.

After translation, in step 5 (\textit{Embedding \& Storage}), the chunks are encoded into embedding vectors in sub-step 5.1 (\textit{Sentence Transformers}\footnote{Python library for generating sentence embeddings based on Transformer models. Available at: \url{https://www.sbert.net/}. Accessed: June 22, 2026.}) using the \texttt{multi-qa-MiniLM-L6-cos-v1} bi-encoder model from the \textit{Sentence Transformers} library. Finally, in sub-step 5.2 (\textit{Vector Database}), the metadata (title and authors), the article pages segmented into chunks, and their respective embeddings are stored in the \textit{PostgreSQL} database, using the \textit{pgvector}\footnote{Open-source extension for vector similarity search in PostgreSQL. Available at: \url{https://github.com/pgvector/pgvector}. Accessed: June 22, 2026.} extension for vector storage.

\subsubsection{Information retrieval}

\begin{figure}[!ht]
    \centering
    \includegraphics[width=0.8\textwidth, trim=0mm 30mm 0mm 20mm, clip]{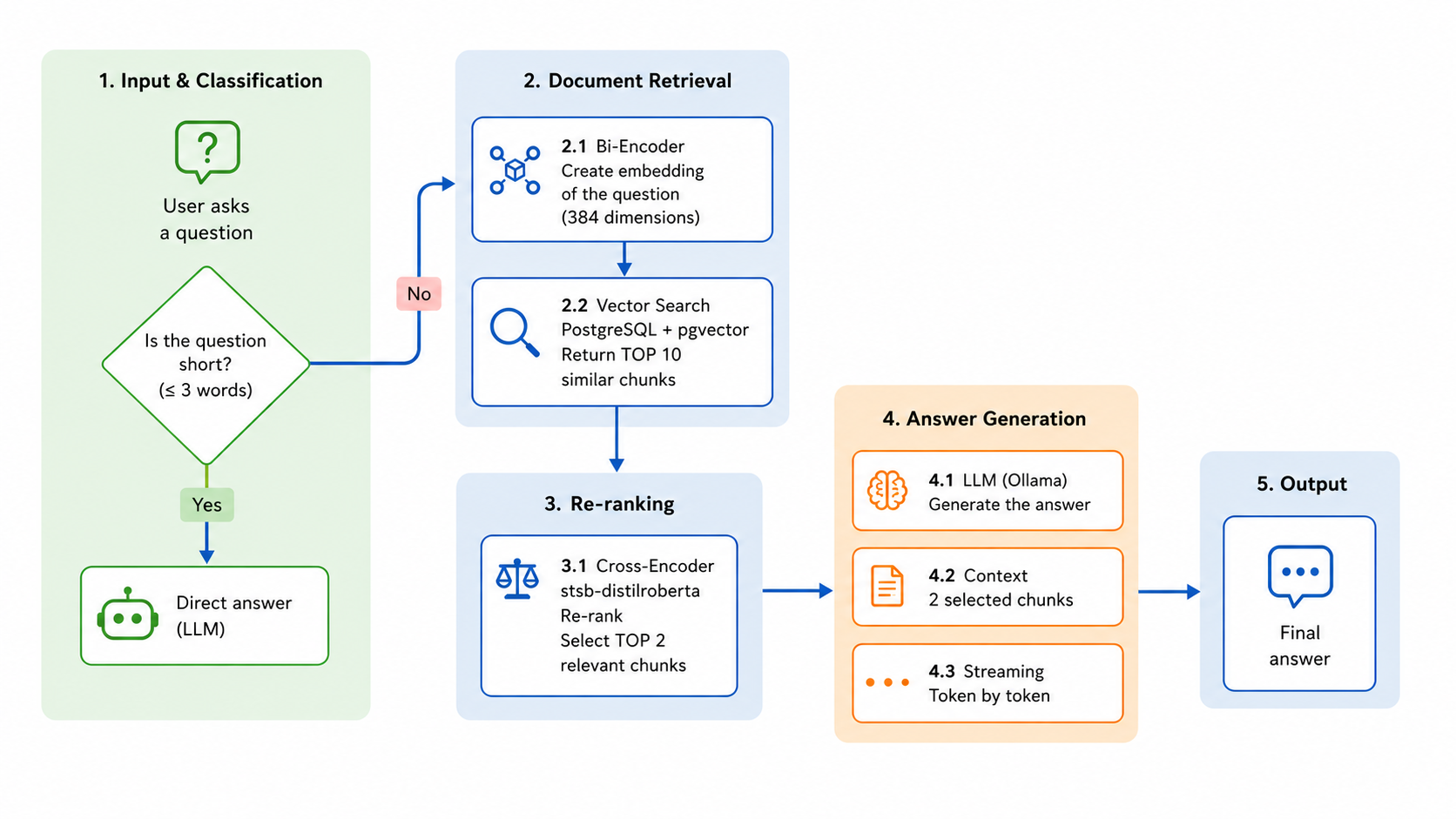}
    \caption{Answer generation pipeline.}
    \label{fig:g1}
\end{figure}

Figure~\ref{fig:g1} presents the information retrieval and answer generation pipeline of the first version of the chatbot.

In step 1 (\textit{Input \& Classification}), upon receiving the user question, the system starts the contextualized retrieval process. First, in step 2 (\textit{Document Retrieval}), the user question is encoded into an embedding vector in sub-step 2.1 (\textit{Bi-Encoder}) using the same bi-encoder (\texttt{multi-qa-MiniLM-L6-cos-v1}). Then, in sub-step 2.2 (\textit{Vector Search}), a similarity search is performed in the database, using Euclidean distance as the distance comparison method. The 10 most similar chunks are retrieved.

In step 3 (\textit{Re-ranking}), the retrieved chunks go through a second re-ranking stage in sub-step 3.1 (\textit{Cross-Encoder}) using the \texttt{cross-encoder/stsb-distilroberta-base} cross-encoder model. This step selects only the two most relevant chunks based on the fine-grained similarity between the user question and each candidate chunk.

With the context retrieved, in step 4 (\textit{Answer Generation}), the system dynamically fills a predefined prompt template, injecting the static behavior instructions, the two re-ranked chunks in sub-step 4.2 (\textit{Context}), and the user question. This properly structured prompt is then sent in sub-step 4.1 (\textit{LLM (Ollama)}) to an LLM executed locally via Ollama, which generates the final answer in Portuguese, streamed via sub-step 4.3 (\textit{Streaming}), resulting in step 5 (\textit{Output}) with the final answer.

\subsection{Overall Architecture, v2}

\begin{figure}[!ht]
    \centering
    \includegraphics[width=0.8\textwidth]{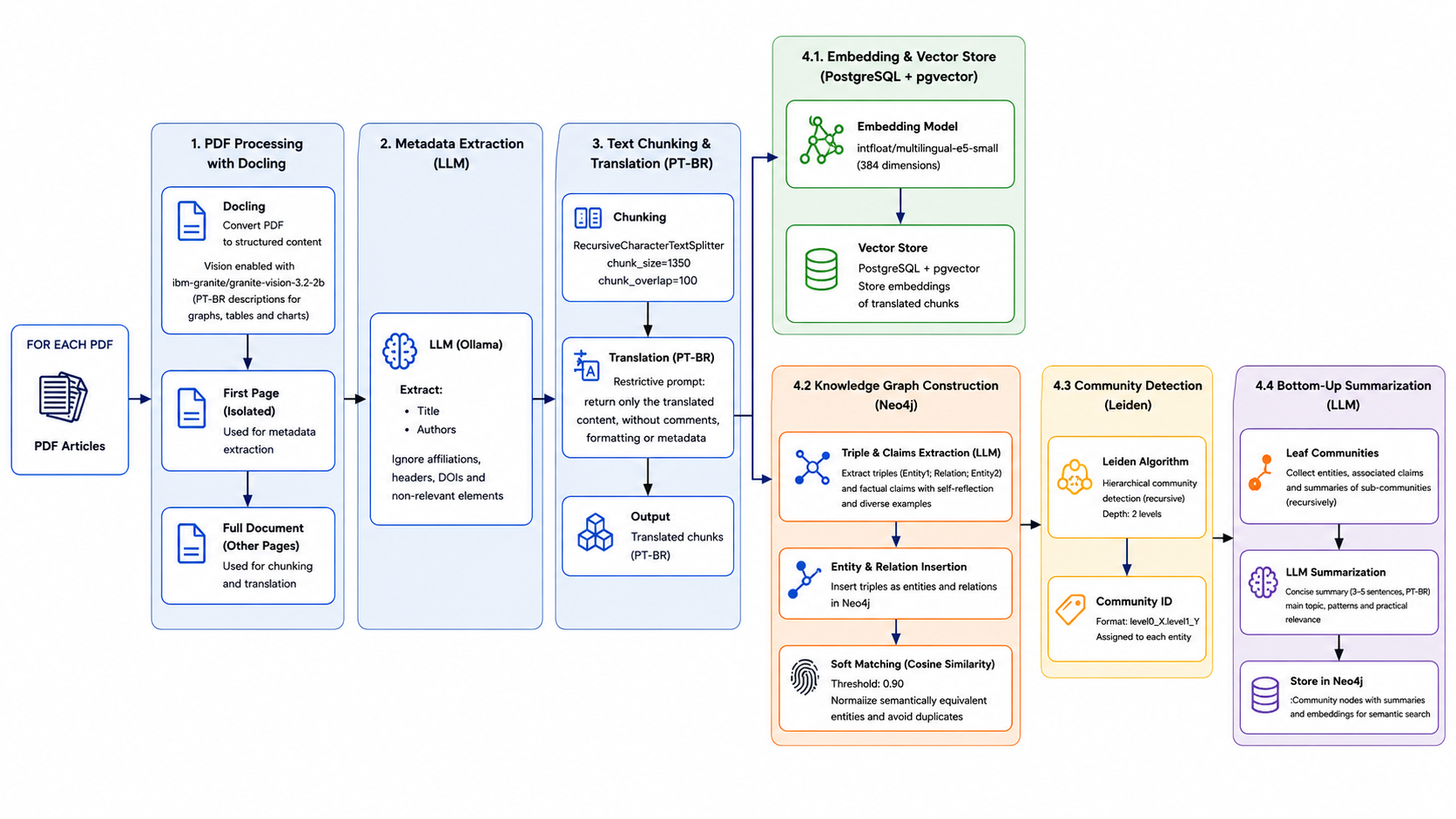}
    \caption{Document ingestion pipeline of the hybrid version.}
    \label{fig:3}
\end{figure}

Figure~\ref{fig:3} presents the ingestion pipeline of the hybrid version of the system. This architecture evolves the previous version by introducing multimedia capabilities in data extraction, a new embedding model, and the parallel construction of a knowledge graph in \textit{Neo4j}\footnote{Graph-oriented database. Available at: \url{https://neo4j.com/}. Accessed: June 22, 2026.}, structured through hierarchical community detection and bottom-up summarization.

The first modification occurs in step 1 (\textit{PDF Processing with Docling}). Unlike the previous version, image description is enabled through the \texttt{ibm-granite/granite-vision-3.2-2b} vision-language model, which generates textual descriptions for charts, figures, and tables, enriching the processed content both on the first page and in the remainder of the document.
In step 2 (\textit{Metadata Extraction (LLM)}), the process of isolating the first page via Ollama is refined so that the LLM explicitly ignores affiliations, headers, DOIs, and other non-relevant elements, retaining strictly the title and the authors.
Step 3 (\textit{Text Chunking \& Translation (PT-BR)}) modifies the segmentation parameters of \texttt{RecursiveCharacterTextSplitter}, increasing the chunk size to 1,350 characters, while keeping the overlap at 100 characters. In addition, the translation of the chunks now uses a restrictive prompt that prevents the LLM from including comments, formatting, or meta-text, ensuring a clean output.
In step 4.1 (\textit{Embedding \& Vector Store (PostgreSQL + pgvector)}), the previous model is replaced by \texttt{intfloat/multilingual-e5-small}, which has a dimension of 384 and offers native support for Brazilian Portuguese for generating the vectors stored in \texttt{PostgreSQL} with the \texttt{pgvector} extension.

The following steps are exclusive to this second architecture. In parallel with the vector storage, each chunk is directed to step 4.2 (\textit{Knowledge Graph Construction (Neo4j)}), starting with the sub-step of \textit{triple \& claims extraction (LLM)}, where a specialized prompt with self-reflection mechanisms and examples extracts triples in the format (Entity1; Relation; Entity2) and claims (factual statements). In the sub-step of \textit{entity \& relation insertion}, the data are populated into \textit{Neo4j}, applying a \textit{soft matching (cosine similarity)} process with a threshold of 0.90 to normalize semantically equivalent entities and avoid duplication.
Over the constructed graph, step 4.3 (\textit{Community Detection}) is applied, which uses the Leiden algorithm recursively for the hierarchical detection of communities at two levels of depth. Each entity receives an identifier in the format \texttt{level0\_X.level1\_Y}, mapping its position in the structure.
Finally, step 4.4 (\textit{Bottom-Up Summarization}) summarizes the communities starting from the leaf communities (the deepest level). The consolidated data are processed in the sub-step of \textit{LLM summarization}, generating concise summaries of three to five sentences in Portuguese about the central theme and the observed patterns. The flow ends in the sub-step of \textit{Store in Neo4j}, where these summaries are saved as \texttt{:Community} nodes, accompanied by their respective vector embeddings to enable semantic search in the query phase.

\subsection{Information retrieval}

\begin{figure}[!ht]
    \centering
    \includegraphics[width=1.0\textwidth]{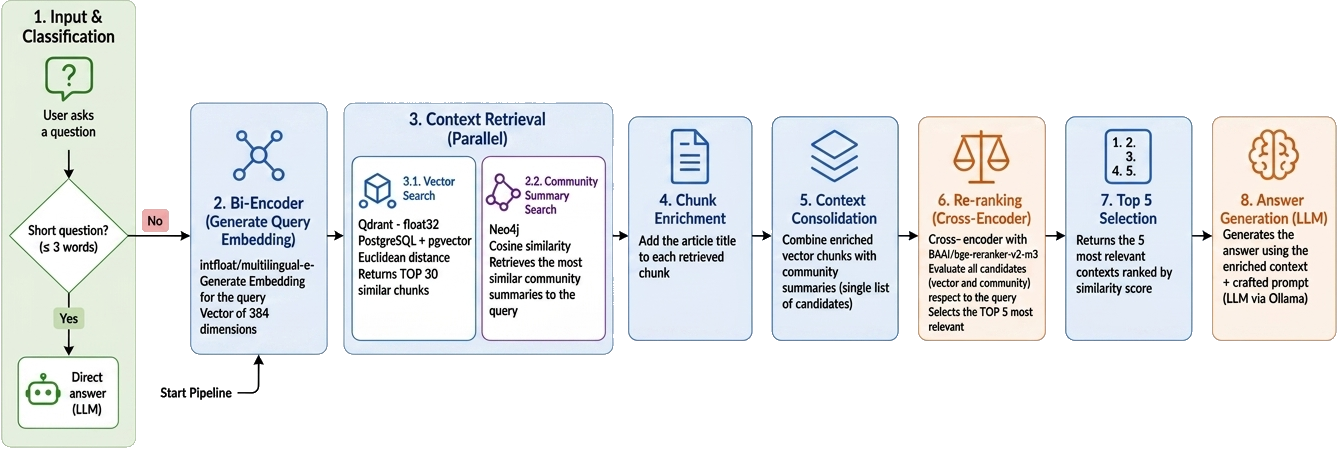}
    \caption{Answer generation pipeline for the hybrid approach.}
    \label{fig:g2}
\end{figure}

Figure~\ref{fig:g2} illustrates the answer generation pipeline of the hybrid version of the system. This architecture evolves the previous model by introducing a parallel retrieval strategy that combines traditional vector search with the search over summaries of communities structured in the graph, in addition to updating the embedding and ranking models.

The flow starts in step 1 (\textit{Input \& Classification}), but changes in step 2 (\textit{Bi-Encoder (Generate Query Embedding)}), where the user question is now encoded by the \texttt{intfloat/multilingual-e5-small} model, replacing the previous embedding model.

The main structural change occurs in step 3 (\textit{Context Retrieval (Parallel)}). While sub-step 3.1 (\textit{Vector Search}) keeps the Euclidean similarity search in the vector database, a concurrent process is triggered in sub-step 3.2 (\textit{Community Summary Search}). In this sub-step, the system performs a vector search based on cosine similarity directly over the community summaries stored in \textit{Neo4j}.
Before the final ranking process, two steps exclusive to this architecture are added. In step 4 (\textit{Chunk Enrichment}), each retrieved chunk receives the metadata of the title of the source article so that the LLM can identify the sources. Then, in step 5 (\textit{Context Consolidation}), the candidates from both sources (vector chunks and graph summaries) are unified into a single list.
The filtering process was also modified in step 6 (\textit{Re-ranking (Cross-Encoder)}), which now uses the \texttt{BAAI/bge-reranker-v2-m3} model. This cross-encoder evaluates the global relevance of the unified list, without distinguishing the origin of the contexts. In step 7 (\textit{Top 5 Selection}), the cut-off criterion is expanded to select the five most relevant contexts, in contrast with the two used in the first version.
Finally, in step 8 (\textit{Answer Generation (LLM)}), this consolidated and enriched block is injected into the final prompt so that the LLM generates the answer in natural language.

\section{Evaluation} \label{sec:metodologia}

The evaluation of the proposed architectures was conducted through a set of 22 questions, each accompanied by its respective expected answer (ground truth). These questions covered the five processed articles. It is important to highlight that these questions were created by the authors of each work themselves, with five questions for the article by Ferreira et al.~\cite{da2013ciclo}, five for Lopes et al.~\cite{lopes2013climatologia}, six for Pontes et al.~\cite{pontes2019role}, three for Cavalcante et al.~\cite{cavalcante2019opposite}, and three for Terassi et al.~\cite{de2023comprehensive}. These articles were chosen because they represent a diversity of topics for the object of study, and also because the authors of these articles are still working as collaborators of ITV at the time of writing of this paper.

The experimental process used the \texttt{gpt-oss-20b} model for inference over the set of questions. The answers were evaluated via the LLM-as-a-judge approach using the DeepEval framework\footnote{\url{https://www.deepeval.com/}}. Seven models acted as judges in the evaluation of the metrics: \texttt{qwen3:32b}, \texttt{qwen3:8b}, \texttt{llama3.3:70b}, \texttt{gemma3:27b}, \texttt{gemma3:12b}, \texttt{mixtral:8x22b}, and \texttt{mixtral:8x7b}.

The generated answers were evaluated based on the following metrics native to the framework: correctness, answer relevance, faithfulness (to the context), contextual relevance, and contextual recall (the ability to retrieve relevant information from the context).
To mitigate the variability inherent to LLM answers and to ensure the statistical robustness of the results, the experiments were executed multiple times, with a repetition frequency between 5 and 10 runs per evaluated scenario. The tests were processed entirely in a dedicated local environment, whose hardware and software specifications comprise: an Intel Core i7-14700KF processor, 64 GB of DDR5 RAM, 1 TB SSD storage, and an NVIDIA GeForce RTX 3090 GPU with 24 GB of VRAM, running under the Ubuntu 24.04.3 LTS operating system. The total estimated time to complete the whole process of data ingestion and judgments was approximately 7 hours.

\section{Results}

\begin{figure}[!ht]
    \centering
    \includegraphics[width=0.7\textwidth]{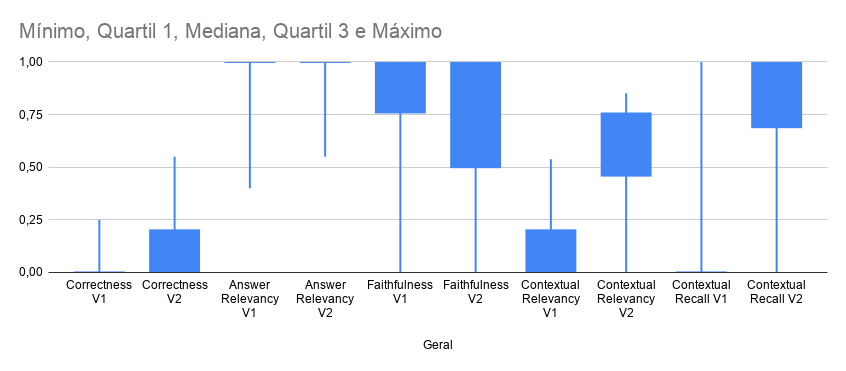}
    \caption{Comparative analysis of the evaluation metrics of the RAG pipeline between versions V1 and V2 through box plots, mapping the statistical distribution of minimum, first quartile ($Q_1$), median, third quartile ($Q_3$), and maximum.}
    \label{fig:p2}
\end{figure}

\begin{figure}[!ht]
\centering
\begin{subfigure}{0.48\textwidth}
    \centering
    \includegraphics[width=\linewidth]{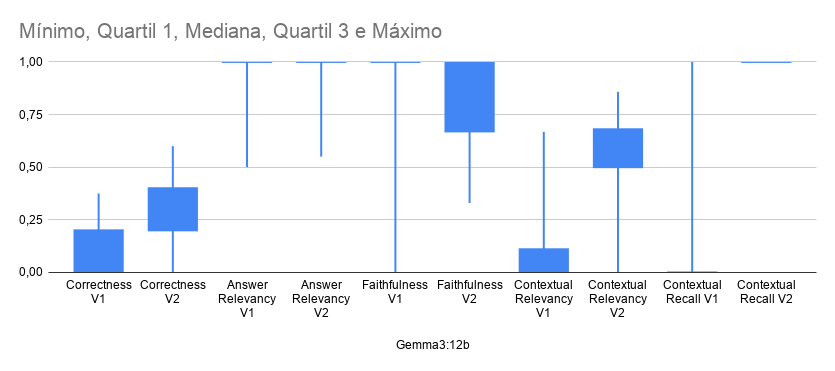}
    \caption{Gemma 3:12b}
    \label{fig:boxplot_gemma3_12b}
\end{subfigure}
\hfill
\begin{subfigure}{0.48\textwidth}
    \centering
    \includegraphics[width=\linewidth]{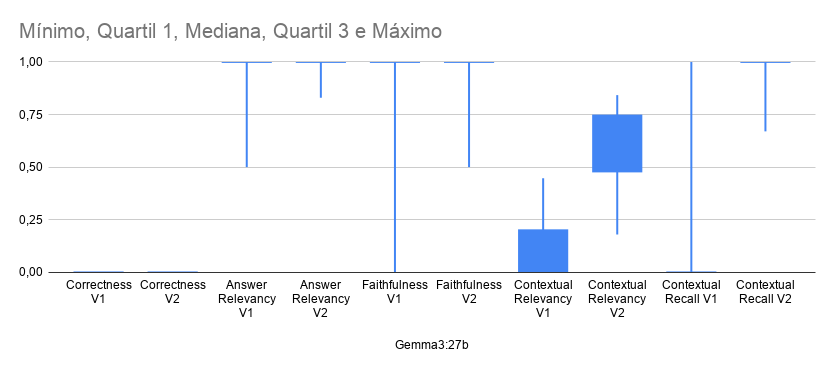}
    \caption{Gemma 3:27b}
    \label{fig:boxplot_gemma3_27b}
\end{subfigure}

\vspace{0.2cm}

\begin{subfigure}{0.48\textwidth}
    \centering
    \includegraphics[width=\linewidth]{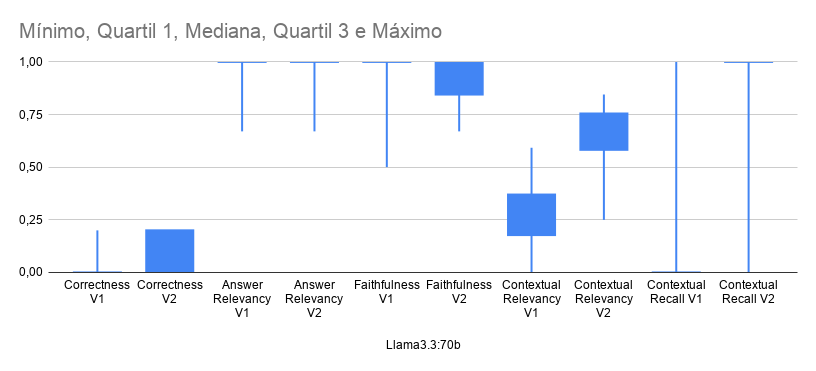}
    \caption{Llama 3.3 70b}
    \label{fig:boxplot_llama3_70b}
\end{subfigure}
\hfill
\begin{subfigure}{0.48\textwidth}
    \centering
    \includegraphics[width=\linewidth]{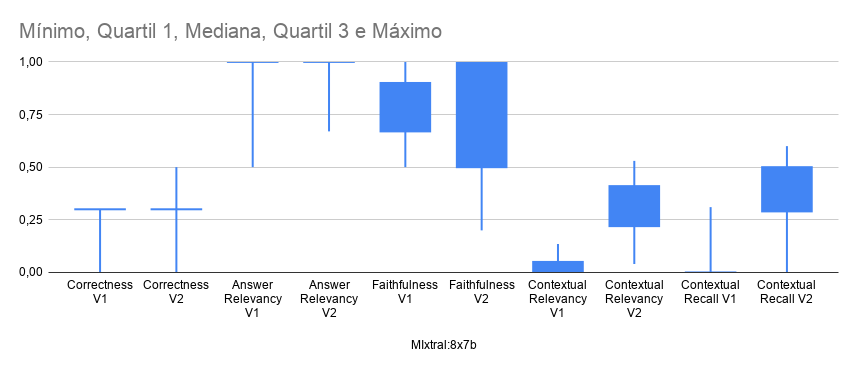}
    \caption{Mixtral 8x7b}
    \label{fig:boxplot_mixtral_8x7b}
\end{subfigure}

\vspace{0.2cm}

\begin{subfigure}{0.48\textwidth}
    \centering
    \includegraphics[width=\linewidth]{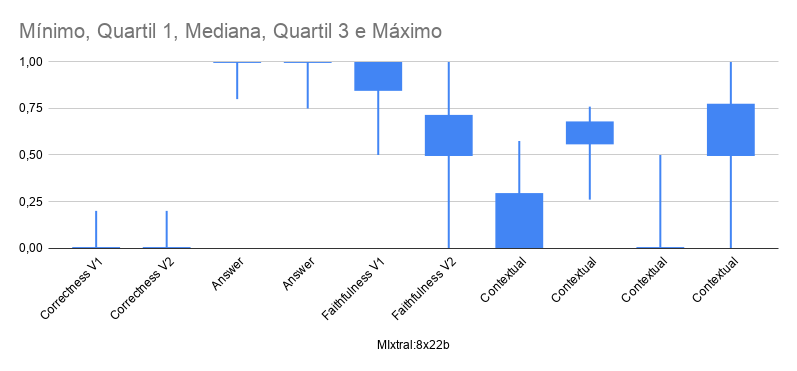}
    \caption{Mixtral 8x22b}
    \label{fig:boxplot_mixtral_8x22b}
\end{subfigure}
\hfill
\begin{subfigure}{0.48\textwidth}
    \centering
    \includegraphics[width=\linewidth]{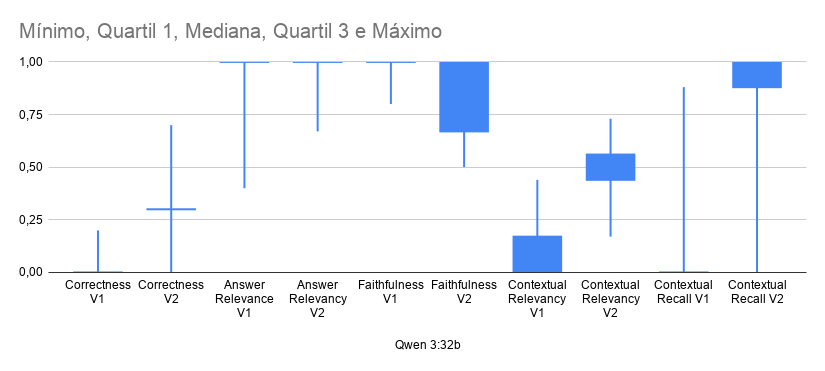}
    \caption{Qwen 3:32b}
    \label{fig:boxplot_qwen3_32b}
\end{subfigure}

\vspace{0.2cm}

\begin{subfigure}{0.48\textwidth}
    \centering
    \includegraphics[width=\linewidth]{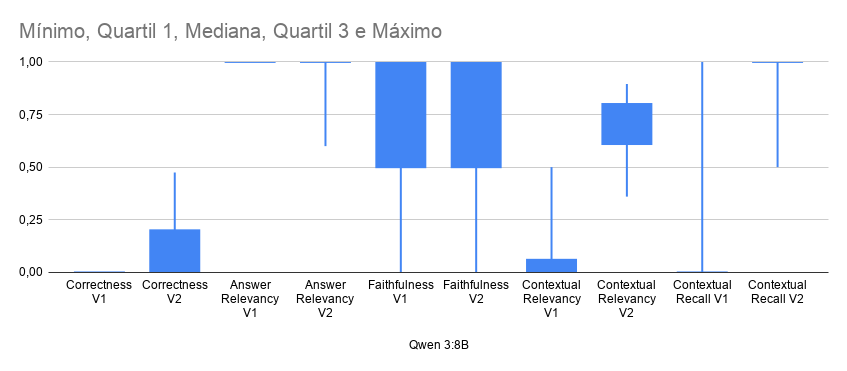}
    \caption{Qwen 3:8b}
    \label{fig:boxplot_qwen3_8b}
\end{subfigure}

\caption{Evaluation box plots for all judge models.}
\label{fig:boxplots_all_models}
\end{figure}

Figure~\ref{fig:p2} shows that version V2 overcame the search bottleneck of V1, presenting an expressive improvement in \textit{Contextual Recall} and \textit{Contextual Relevancy}. This refinement reduced the omission of knowledge and raised the accuracy of the generation (\textit{Correctness}), while keeping the \textit{Faithfulness} and \textit{Answer Relevance} scores stable. Additionally, Figure~\ref{fig:boxplots_all_models} details the variability and the consistency of the seven judge models through box plots, allowing a more robust diagnostic analysis of the proposed architectures.

\paragraph{Aggregate Analysis: Robustness and Reliability.}
The aggregate results confirm substantial gains of the hybrid model, but an equally important finding emerges from the detailed analysis: the \textbf{variability of the evaluations decreases drastically} in the transition from V1 to V2. While Vector RAG presents box plots with wide spreads (indicating disagreement among judges), the
hybrid model exhibits compressed box plots, especially for \textit{contextual recall}, where all seven models agree on medians between 0.75 and 0.95. This indicates that GraphRAG not only improves average performance, but also produces more \textbf{predictable and reliable} retrievals, a critical aspect for production
systems. \textit{Contextual relevance} follows a similar pattern: V1 ranges from 0.20 to 0.35, while V2 concentrates between 0.60 and 0.70 for almost all judges.

\paragraph{Divergences Among Judge Models.}
The individual analysis of the models reveals heterogeneity in the evaluations that provides clues about the characteristics of the architectures. \textbf{Gemma 3:12b} (Figure~\ref{fig:boxplots_all_models}-(a)) evaluates with particularly strict criteria, presenting some of the tightest box plots of the batch, yet its medians for \textit{contextual recall} in V2 reach approximately 0.95---the highest among all. This suggests that smaller models, when
calibrated, can be more discriminative. In contrast, \textbf{Llama 3.3 70b} (Figure~\ref{fig:boxplots_all_models}-(c)) exposes a critical weakness of vector RAG: its box plot of \textit{contextual recall} in V1 presents a high outlier ($\sim$0.6) accompanied by a low median ($\sim$0.15), indicating that the larger model occasionally captures adequate retrievals, but fails systematically. Crucially, V2 \textbf{eliminates this variability}, with a median of $\sim$0.90 and minimal spread, suggesting that GraphRAG stabilizes retrieval even under different types of questions.
\textbf{Mixtral 8x22b} (Figure~\ref{fig:boxplots_all_models}-(e)) stands out for exhibiting the largest variability in \textit{faithfulness}, with a spread of approximately 0.30 even in V2 (ranging from 0.6 to 1.0). While other models agree that \textit{faithfulness} drops moderately from 0.90 to 0.80, Mixtral detects larger inconsistencies. Interpretation: when expanding from 2 to 5 chunks, some answers introduce nuances that may be only partially grounded in the context, revealing a trade-off between contextual completeness and strict fidelity---a characteristic of the design that deserves attention.

\paragraph{The Correctness Paradox and the Validation of the Gains.}
A pattern emerges when observing the \textit{correctness} metric in the box plots: all seven judge models agree that this metric remains problematic (medians below 0.40 even in V2), despite the improvement in context retrieval. \textbf{Qwen 3:32b} (Figure~\ref{fig:boxplots_all_models}-(f)) exemplifies this validation: its progression is uniform and balanced (both for V1 and V2), reflecting the aggregate gains without distortions. Its box plots of \textit{answer relevance} remain extremely tight at 0.98 for both versions, evidencing that the improvement is robust, and not an artifact of one specific evaluator.
These results show that the hybrid approach, by integrating vector search, the knowledge graph, and re-ranking, tends to significantly improve context retrieval, and also \textbf{reduces inter-model variability} and increases the predictability of the evaluations. However, the persistence of low \textit{correctness} suggests that additional post-generation factual verification mechanisms are needed for systems that demand high factual fidelity; a possible direction for future work.

\section{Limitations}

The generalizability of the results obtained is limited by the small size and the highly specific scope of the dataset, which has only 22 questions restricted to the domain of Amazon climatology, in addition to being subject to the potential bias arising from its manual selection. Additionally, the adoption of the LLM-as-a-judge approach introduces a direct dependency on the quality of the evaluator models and on their respective inherent biases, which makes the inclusion of a validation by domain experts advisable in future work. Finally, the execution of the experiments in a local and controlled environment via Ollama, together with the purely empirical tuning of the hyperparameters, imposes computational and operational restrictions, requiring additional validations before these results can be extrapolated to production scenarios or applied to new contexts.

\section{Related Work}
\label{sec:trabalhos_relacionados}

The work of Han et al.~\cite{han2025rag} provides the most direct comparison with our approach by systematically evaluating vector RAG and GraphRAG variants in QA tasks (single and multi-hop) and summarization in the domains of Wikipedia, news, and literature. The authors concluded that the technologies are complementary: RAG stands out in local factual queries, while GraphRAG is superior in complex reasoning. In contrast with that focus on general benchmarks, the present paper applies a hybrid architecture to the climatology domain, validating the use of Leiden community detection and semantic triples to integrate dispersed scientific data.

Regarding infrastructures and metrics, three studies propose methods that either contrast with or support the choices of this research. Laban et al.~\cite{laban2407summary} introduced the SummHay benchmark to evaluate the summarization of multiple long documents, revealing that advanced RAG systems still fall short of humans in insight coverage. This scenario corroborates our choice of bottom-up community summarization to synthesize multiple articles.

Pradeep et al.~\cite{pradeep2025great} proposed AutoNuggetizer, which automates evaluation based on atomic facts (nuggets), demonstrating a strong correlation between the judgment of LLMs and that of human experts in the TREC 2024 RAG Track. This atomic factual precision emerges as a future alternative to refine the correctness metric used in our setting. 
Finally, in the scenario of optimization in specific domains, Setty et al.~\cite{setty2024improving} evaluated the impact of zero-shot techniques (such as query expansion and re-ranking) via the Ragas framework in the financial domain. The results indicate that re-ranking mitigates the limitations of traditional dense retrievers in highly complex texts. This conclusion validates the use of cross-encoders (such as the BGE-reranker) in the hybrid architecture proposed in this paper.

\section{Concluding Remarks}\label{sec:conclusao}

This paper compared classical vector RAG with a hybrid approach based on GraphRAG. Validated with 22 complex questions and multiple LLMs, the hybrid architecture outperformed the traditional one with gains of 160\% in contextual relevance and 177\% in contextual recall, integrating local details with global structures of the graph and overcoming document isolation.


\medskip

\bibliography{referencias}

\end{document}